\documentclass[aps,prl,twocolumn,superscriptaddress,showpacs]{revtex4-1}
\usepackage{graphicx}
\usepackage{bm}
\usepackage{color}
\usepackage{hyperref}

\newcommand{\ave}[1]{{\langle #1\rangle}}

\begin{document}
\title{Breakdown of the generalized Gibbs ensemble
for current--generating quenches}
\author {Marcin Mierzejewski}
\affiliation{Institute of Physics, University of Silesia, 40-007 Katowice, Poland}
\author{Peter Prelov\v sek}
\affiliation{Faculty of Mathematics and Physics, University of Ljubljana, SI-1000 Ljubljana, Slovenia }
\affiliation{J. Stefan Institute, SI-1000 Ljubljana, Slovenia }
\author{Toma\v z Prosen}
\affiliation{Faculty of Mathematics and Physics, University of Ljubljana, SI-1000 Ljubljana, Slovenia }
  
\begin{abstract}
We establish a relation between two hallmarks of integrable systems: the relaxation
towards the generalized Gibbs ensemble (GGE) and the dissipationless charge transport.
We show that the former one is possible only if the so called Mazur bound on the
charge stiffness is saturated by local conserved quantities. As an example 
we show how a non--GGE steady state with a current can be generated in the one-dimensional 
model of interacting spinless  fermions with a flux quench. Moreover an extended GGE 
involving the quasi-local conserved quantities can  be formulated for this case.        
\end{abstract}
\pacs{72.10.-d,75.10.Pq, 05.60.Gg,05.70.Ln}

\maketitle

Recent advances in experiments on ultracold atoms  together with new computational techniques 
have significantly broadened our understanding of relaxation processes in closed many-body quantum systems. 
It is commonly accepted that in generic macroscopic systems the long--time averages of  {\em local} observables  
coincide with the results for the statistical Gibbs ensemble \cite{Goldstein2006,Linden2009,Riera2012,polkrev} 
and are uniquely determined by few parameters related to conserved quantities, in particular the system's energy 
and particle number.  Due to the presence of macroscopic number of conserved quantities such a simple scenario 
is not applicable to integrable systems \cite{Manmana2007,santos2011,our2013}. However, there is a  large and 
still growing evidence that relaxation in the latter systems is consistent with the generalized Gibbs ensemble (GGE)  \cite{gge,Eckstein2012,Cassidy2011,Gogolin2011,essler2014}, where the density matrix  is determined not only by the 
Hamiltonian $H$ and particle number $N$ but also by other {\em local} conserved quantities $Q_i$, i.e., 
$\rho_{GGE} \sim \exp\left[-\beta (H-\mu N)-\sum_i \lambda_i Q_i\right]$. 

In this Letter we focus on the relaxation dynamics of one of the most studied integrable models: the 
model of interacting spinless fermions, being equivalent to the anisotropic 
Heisenberg ($XXZ$) model for which the set of $Q_i$ has been established \cite{tetelman,grabowski}. 
We show that $\rho_{GGE}$ as generated only by {\em local} integrals of motion  $Q_i$ doesn't exhaust all generic
stationary states in the metallic (easy plane) regime. Instead, there are cases for which one should lift the requirement 
of locality of the conserved quantities and allow also for quasi--local integrals of motion 
\cite{tomaz_quasilocal11,tomaz_quasilocal13}. In this Letter we call them non--GGE states, 
however we stress that these states can be viewed also as
``extended GGE'',  where the extension concerns the locality of operators. 
Such operators have the parity opposite to local ones $Q_i$. We identify one of such quasi--local quantities 
as  the time--averaged particle current operator and we construct as well as verify it explicitly.

It has been well recognized that integrable systems in spite of interaction reveal
anomalous transport properties at finite inverse temperatures $\beta=1/T$, e.g. the  dissipationless particle current. 
This property is manifested by a nonvanishing  charge stiffness $D(\beta <\infty )$ \cite{zotos1996,zotos1997,robin2013}, 
which in turn is bounded from below by the local conservation laws via the Mazur bound \cite{mazur,zotos1997}.    
The dissipationless transport and the relaxation towards GGE are probably the most
prominent hallmarks of integrability, still they have been studied independently of each other so far. 
While it has been clear that in certain regimes the standard Mazur bound with only local $Q_i$ 
does not exhaust the phenomenon of  dissipationless transport and 
$D(\beta < \infty )>0$ \cite{zotos1997}  we show in this Letter that GGE should be extended 
by taking into account  quasi--local conserved quantities of different parity,
in particular the time averaged current, which  saturate $D(\beta \to \infty )$  
within the  Mazur bound.   	       

We study a prototype one-dimensional (1D) model of interacting particles, the tight-binding model of 
spinless fermions on $L$ sites at half filling (with $N=L/2$ particles) and with periodic boundary conditions 
\cite{ring1,ring2,ring3,ring4},
\begin{equation}
H(t)= -t_h\sum_{j=1}^L ({\mathrm e}^{i \phi(t)}\; c^{\dagger}_{j+1}c_j + {\rm h.c.}) + V \sum_{j=1}^L 
\tilde{n}_j \tilde{n}_{j+1},
\label{ham}
\end{equation}
where $n_j= c^{\dagger}_{j}c_j$, $\tilde n_j=n_j -1/2$, $t_h$ is the hopping integral and  $V$ is the repulsive 
interaction on nearest neighbors.  
The model (\ref{ham}) is equivalent to the anisotropic Heisenberg ($XXZ$) model with the exchange interaction 
$J = 2t_h$ and the anisotropy parameter  $\Delta= V/2t_h$. However, we stay within the fermionic representation, 
where the phase $\phi(t)$ has a clear physical meaning: it represents the 
time-dependent magnetic flux which induces the electric field  $F(t)=-\partial_t \phi(t)$. Further on we use 
$\hbar=k_B=1$
and units in which  $t_h=1$.  We consider here the metallic (easy--plane) regime  $V < 2$ ($\Delta <1$) where 
the system exhibits a 
ballistic particle (spin) transport at $T>0$  \cite{zotos1996,zotos1997, zotos1999,*benz,review2007,
shastry, herbrych2011, Marko2011,Sirker2009,robin2013,Tomaz2011,my3}. 

The charge  stiffness  $D(T)>0$ has been introduced via the $T>0$ generalization of the Kohn's \cite{kohn} 
argument of the  level curvatures $\epsilon_n(\phi)$ \cite{castella1995,zotos1996}. 
It is still a challenging problem since it cannot be derived from 
{\em local} conservation  laws \cite{zotos1997,zotos_unpub}. To  explore this relation 
we study  in the following the standard particle current 
$J= \sum_j (i {\mathrm e}^{i \phi(t)}\; c^{\dagger}_{j+1}c_j +{\rm h.c.})$ as well a less common current 
with a correlated hopping to next-nearest neighbors $J'= \sum_j (i {\mathrm e}^{2i \phi(t)}\; c^{\dagger}_{j+2}
\tilde n_{j+1} c_j +{\rm h.c.})$. The central point in our reasoning is the particle--hole (parity) transformation 
\begin{equation}
c_{i} \rightarrow (-1)^{i}  c^{\dagger}_{i}, \label{sym}
\end{equation}
which (for $\phi=0$) does not alter the Hamiltonian $H \rightarrow H$ (at half filling) nor 
the local conserved quantities $Q_i \rightarrow Q_i$  \cite{zotos1997} but reverses 
the currents $J  \rightarrow -J$ and $J'  \rightarrow -J'$, hence $J(J')$ and $Q_i$ have different parities.    

\begin{figure}
\includegraphics[width=0.5\textwidth]{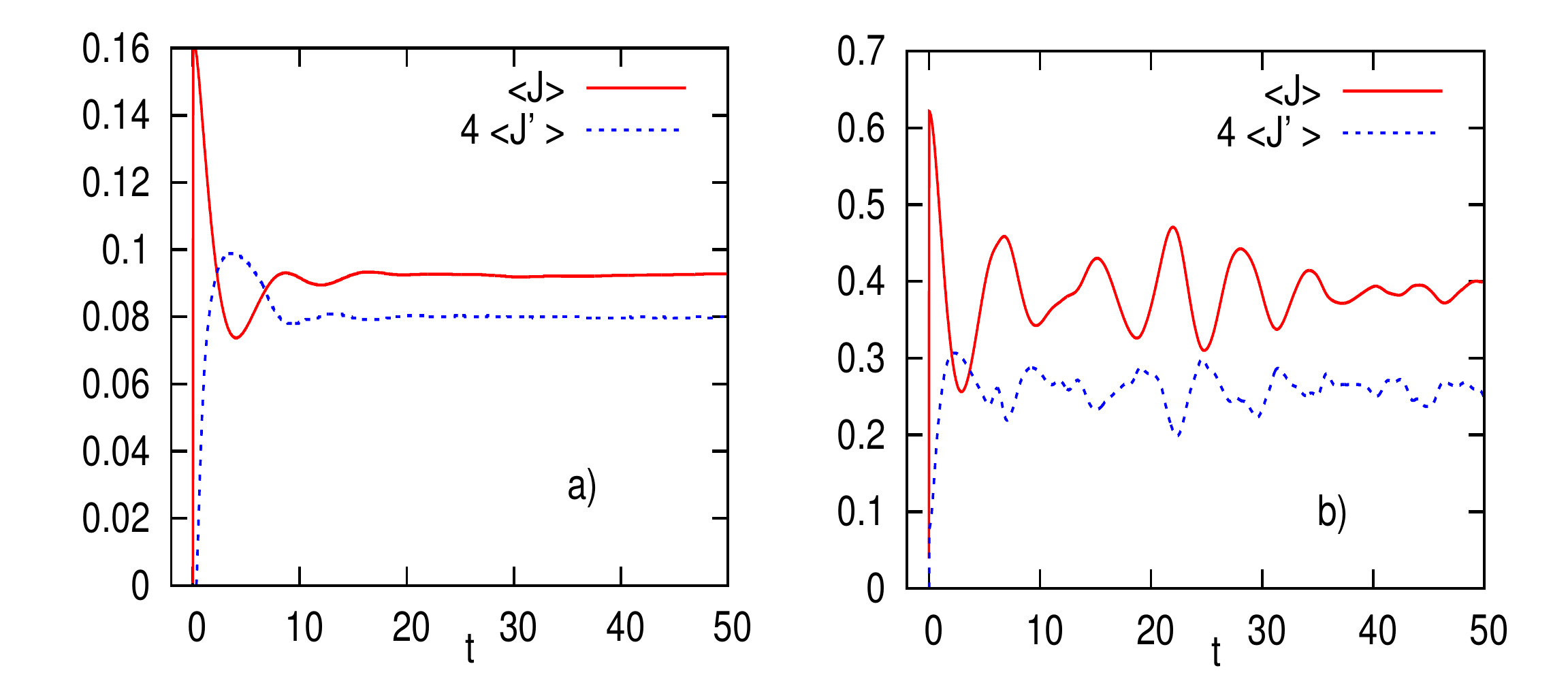}
\caption{
(Color online) Time--dependence of $\langle J \rangle$  and 
$4 \langle J' \rangle$ after quenching the flux  at $t=0$ for $L=26$ and $V=1$.  a) The  system is 
initially  in equilibrium  microcanonical state with $\beta=0.35$ while after the quench it has effective 
$\beta=0.15$. b) The system is in the ground state and the flux is quenched from  $\phi_0=\pi/2$ to $0$. }
\label{fig1}
\end{figure}

We start with numerical studies of a quantum quench which generates a non--GGE  steady state. 
We consider a system which for $t<0$ is either in the ground state or in the equilibrium canonical or microcanonical
state \cite{mclm}. In the latter case we generate a state $ \rho(0)=|\Psi(0) \rangle \langle \Psi(0) | $ 
for the target energy $E_0= \langle \Psi(0)| H(0) |\Psi(0) \rangle$ and with a small energy 
uncertainty $\delta^2 E_0=\langle \Psi(0)| [H(0)-E_0]^2|\Psi(0) \rangle $ as discussed in Refs. \cite{my1,my2}. 
The time--evolution shown in Fig. \ref{fig1}  has been obtained by the Lanczos propagation 
method \cite{lantime,my1,my2}. 

At $t=0$ the magnetic flux is suddenly decreased from the initial value $\phi(0)= \phi_0>0$ to $\phi(t>0)=0$.  
Such a quench is equivalent to a pulse of the electric field $F(t)=\phi_0 \delta(t)$ hence it generates 
the particle current $ \ne 0$. As shown in   Fig. \ref{fig1} this quench induces also  
$\langle J'(t) \rangle \ne 0$, 
however the latter quantity increases gradually in contrast to the  instantaneous  generation of $\langle J(t>0) \rangle$.  
Both currents reach for $t \to \infty$ finite steady 
values, clearly visible in Fig. \ref{fig1}, being the signature of dissipationless transport.  
Still the residual values $\langle J \rangle \ne 0$ and  $\langle J' \rangle \ne 0$ cannot be 
explained within the GGE--scenario since  $\mathrm{ Tr}\{\rho_{GGE} J\}= \mathrm{Tr}\{\rho_{GGE} J'\}=0 $ 
due to different symmetries under particle--hole transformation at half filling \cite{zotos1997}. 


The first objective of this Letter is to establish the symmetry decomposed time averaged density matrix
\begin{equation}
\bar{\rho}=\lim_{\tau \rightarrow \infty} \frac{1}{\tau} \int_0^\tau dt \rho(t)
=\bar{\rho}_e+\bar{\rho}_o,
\label{barrho} 
\end{equation}
where $\bar{\rho}_o$ and $\bar{\rho}_e$ are odd and even under the transformation (\ref{sym}), respectively. 
Since $\mathrm{ Tr} \{ \bar{\rho} J \}= \mathrm{ Tr} \{ \bar{\rho}_o J \}$ the odd component of
the density matrix $\bar{\rho}_o$ is essential for the nonvanishing current $\langle J(t>0) \rangle$, 
while this component is missing in 
$\rho_{GGE}$.  At this stage it is instructive to recall the linear--response (LR) results 
\begin{equation}
\langle J(t) \rangle=L\int_{-\infty}^{\infty} \mathrm{ d}{\omega}
e^{-i \omega t} F(\omega) \sigma(\omega), 
\end{equation}
where the optical conductivity $\sigma(\omega)$ consists of the regular and the ballistic parts with the latter one 
determined  by the charge stiffness $D$: $\sigma_{bal}(\omega)=2Di/(\omega+i0^{+})$. The quench of 
flux induces an electric field $F(\omega)=\phi_0/(2\pi)$ and the regular (dissipative) part of conductivity 
becomes irrelevant in the long--time regime. Then we get within the LR, i.e. for $\phi_0 \ll 1$,
\begin{equation}
\lim_{t\to\infty}\langle J(t) \rangle= 2 L D \phi_0.
\label{lr}
\end{equation}
 An important message following from LR, Eq.(\ref{lr}), is that the non--GGE component of the density matrix 
has to contain contributions which are linear in $\phi_0$ and, therefore, can be singled out already 
within the first--order perturbation expansion in $\phi_0$. The unperturbed Hamiltonian $H_0=H(t < 0 )$ is given by 
Eq. (\ref{ham}) with $\phi(t)$ replaced by $\phi_0$, while the perturbation reads 
$H'(t)=H(t)-H_0=(\phi_0-\phi(t))J_0$, where $J_0=J(t<0)$, so that $H'(t>0)=\phi_0 J_0$. 
For the sake of clarity all quantities obtained 
with the flux $\phi_0$ will be marked with a label "0", in particular
the eigenvalues $E_{m0}$ and the eigenvectors $|m0\rangle$ of $H_0$. 	
The degeneracy of energy levels plays an important role and  should not be neglected. 
Hence, we diagonalize  the current operator in each subspace spanned by degenerate eigenstates and 
take the eigenvectors of $J$ as the basis vectors of this subspace, i.e.,
$\langle m|J|n\rangle \propto \delta_{mn}$ if $E_m=E_n$ (within a subspace only). 


We assume that the system is initially in a thermal sate, 
i.e. $\rho_0=\sum_m p_{m0} |m0 \rangle \langle m0|$ with 
$p_{m0}=\exp(-\beta E_{m0})/Z_0 $. Then, in the Schr\"odinger picture
one obtains
\begin{eqnarray}
&&\rho(t>0) = \sum_m p_{m0} e^{-iH_0 t} U(t)   |m0 \rangle \langle m0| U^{\dagger}(t) e^{iH_0 t}, \nonumber \\
&&U(t>0)=T_{t} \exp \left( -i \int_0^{t} \mathrm{d} t'
H'_{I}(t') \right), \label{rhot}
\end{eqnarray}
where $H'_{I}(t')$ is the perturbation in the interaction picture. 
Our aim is to explicitly express $\bar \rho$ within the LR to the quench $\phi_0$.
A straightforward calculation of Eqs.~(\ref{barrho}),(\ref{rhot}) to first order in $\phi_0$ yields
\begin{equation}
\bar{\rho} \simeq \rho_0+ \phi_0 \sum_{E_{m0}\ne E_{n0}}  
\frac{p_{n0}-p_{m0}}{E_{n0}-E_{m0}} \langle m0|J_0|n0\rangle |m0 \rangle \langle n0|.
\label{rho1}
\end{equation}
We should also take into account  the change of current operator due to flux, hence
\begin{equation}
J=J(t>0)=J_0-\phi_0 H^k_0, \label{j1}
\end{equation}
where $H^k_0$ is the kinetic part of $H_0$, Eq.~(\ref{ham}).  
Using Eqs. (\ref{rho1}),(\ref{j1}),(\ref{lr}) one then restores  the LR result
for the equilibrium charge stiffness \cite{castella1995,zotos1996}
\begin{equation}
D=\frac{1}{2L} \left[- \langle H^k_0\rangle + \sum_{E_{m0}\ne E_{n0}}  
\frac{p_{m0}-p_{n0}}{E_{m0}-E_{n0}} |\langle m0|J_0|n0\rangle |^2
\right].
\end{equation} 
Eq. (\ref{rho1}) does not yet accomplish our aim of decomposing 
$\bar{\rho}$ into odd and even parts with respect to
(\ref{sym}) after the quench $\phi(t>0) = 0$. We achieve this by using
again the first--order perturbation theory for $H_0=H-\phi_0 J $ and 
$J_0=J+\phi_0 H^k$, where now $H,H^k$ and $J$ are the operators after the quench, 
i.e. at $\phi=0$. Substituting 
\begin{eqnarray}
E_{n0}&=& E_n-\phi_0 \langle n|J| n\rangle, \nonumber \\ 
|n0 \rangle &=& |n \rangle- \phi_0 \sum_{m:E_m \ne E_n}
\frac{\langle m | J | n  \rangle}{E_n-E_m} |m \rangle, 
\end{eqnarray} 
into Eq. (\ref{rho1}), and assuming that there is no particle current in the initial thermal state, 
we finally obtain
\begin{equation}
\bar{\rho}=\sum_n p_n  | n \rangle \langle n| \left(1+\beta \phi_0 \bar{J} \right), 
\label{rho2}
\end{equation}
where $\bar{J}$ is the time-averaged steady--current operator
\begin{equation}
\bar{J}=\lim_{\tau \rightarrow \infty} \frac{1}{\tau} \int_0^\tau dt 
e^{iHt} J e^{-iHt}=\sum_n \langle n | J | n  \rangle 
|n  \rangle \langle n |.
\label{jbar}
\end{equation}
The LR results [Eq. (\ref{lr})] is immediately restored, however this time with the 
alternative form of the charge stiffness but equivalent for $\beta<\infty$ and in the thermodynamic 
limit \cite{zotos1997}
\begin{equation}
D=\frac{\beta}{2L } \sum_n p_n \langle n | J | n  \rangle^2.  
\end{equation}
By definition $\bar{J}$ is an integral of motion $[H,\bar{J}]=0$. It is important to note that
$\mathrm{Tr} \bar{J}^2 /{\cal N} \propto L$ where ${\cal N}={\mathrm{Tr}}\,1$ is the dimension of the Hilbert space,
already implies that  $\bar{J}$ is a quasi--local quantity.
Since at $\beta \to 0$,
\begin{equation}
\frac{1}{{\cal N}} \mathrm{Tr} \bar{J}^2  =2L \tilde{D},\quad {\rm where}\quad
\tilde{D}=\lim_{\beta \to 0} D(\beta)/\beta,
\label{druid}
\end{equation} 
the quasi-local character of $\bar{J}$ is consistent with the well established fact 
that the charge stiffness is an intensive quantity. 

We now turn to the question  of whether $\bar{\rho}$ is compatible with
$\rho_{GGE}$ and the answer is clearly negative.
 A necessary and a sufficient condition for such compatibility, to leading order in the quench $\phi_0$, 
would be a decomposition in terms of local conserved $Q_i$,
\begin{equation}
\bar{J}=\sum_i \alpha_i Q_i, \label{test}
\end{equation} 
holding for some set of $\alpha_i$. 
Assuming that  $\mathrm{Tr} \{Q_i Q_j \} \propto \delta_{ij}$ we can employ the inequality
\begin{equation}
\mathrm{Tr} \{ (\bar{J} - \sum_i a_i Q_i)^2 \} \ge 0, 
\label{mini}
\end{equation}
which holds for any $a_i$ and becomes an equality only for the GGE state with $a_i=\alpha_i$. 
Now we can follow original steps by Mazur \cite{mazur}.  We minimize the lhs of Eq. (\ref{mini}) 
with respect to $a_i$, 
\begin{equation}
a_i=\frac{\mathrm{Tr}\{\bar{J} Q_i\}}{\mathrm{Tr}\{ Q^2_i\} }
=\frac{\mathrm{Tr}\{J Q_i\}}{\mathrm{Tr}\{ Q^2_i\} },
\end{equation}  
and substitute this result in (\ref{mini}) to obtain the
Mazur inequality for $\beta \to 0$
\begin{equation}
\mathrm{Tr}\{ \bar{J}^2 \}  \ge 
\sum_i \frac{\mathrm{Tr}\{J Q_i\}^2}{\mathrm{Tr}\{ Q^2_i\} },
\label{mazur}
\end{equation}
which is the Mazur bound on charge stiffness at $T \to \infty$ [see Eq. (\ref{druid})]. Since  this inequality 
turns into equality for GGE states, so should the Mazur bound. In other words
relaxation towards GGE is possible provided the Mazur bound 
saturates the charge stiffness.This relation 
holds for an arbitrary filling $N/L$. In particular for  $N/L=1/2$ one finds $\mathrm{Tr}(Q_i J)=0$ due to the symmetry (\ref{sym}), hence the rhs of (\ref{mazur}) vanishes, 
and our quenched dynamics does {\em not} relax to GGE.

As has been shown in Refs. \cite{tomaz_quasilocal11,tomaz_quasilocal13}, another set of non-local, 
but {\em quasi-local} conserved Hermitian operators $\{ Q(\varphi) \}$ exists for a dense set commensurate interactions
$\Delta=\cos(\pi l/m)$, with $l,m$ integers, densely covering the range $|V| < 2$. They are all odd under (\ref{sym}), 
$Q(\varphi)\to -Q(\varphi)$. Quasi-locality implies linear extensivity $\mathrm{Tr}\{Q(\varphi)^2\}/{\cal N} \propto L$, 
similarly as for the local conserved operators $Q_i$,
while $\mathrm{Tr}\{J Q(\varphi)\} /(L{\cal N})= {\rm const}$, making them suitable for 
implementing the Mazur bound.
For $\Delta=\cos(\pi/m)$ for which $T\to\infty$ limit of the Bethe ansatz result \cite{zotos1999,*benz} is available it has been shown 
\cite{tomaz_quasilocal13} to agree precisely with the Mazur bound, so one may conjecture that the latter is now indeed saturated.
Hence our argument (\ref{test}-\ref{mazur}) can be used to argue that the complete time-averaged current can be expressed in terms of an integral
\begin{equation}
\bar{J} = \int_{{\cal D}_m} d^2\varphi f(\varphi) Q(\varphi)
\end{equation}
where $f(\varphi) = c_m/|\sin\varphi|^{4}$ for a suitable constant $c_m$ (see \cite{tomaz_quasilocal13}) 
and ${\cal D}_m$ is a vertical strip in the complex plane with $|{\rm Re}\varphi-\pi/2|<\pi/(2m)$.
After straightforward calculation, again using the notation and machinery of \cite{tomaz_quasilocal13}, 
one arrives at the explicit {\em matrix-product} expression for $\bar{J} = i(J_+- J^\dagger_+)$ in terms of local operators
\begin{equation}
J_+ = \sum_j \sum_{r\ge 2} J^{(r)}_j
\end{equation}
with 
\begin{eqnarray}
&&J^{(r)}_1 = \!\!\!\!\!\sum_{s_2,\ldots,s_{r-1}}^{\{0,z,\pm\}}\!\!\!\!g_{s_2\ldots s_{r-1}} (B^{s_2}\cdots 
B^{s_{r-1}})_{11} \sigma^-_1 \sigma^{s_2}_{2} \cdots \sigma^{s_{r-1}}_{r-1}\sigma^{+}_{r}, \nonumber \\ 
&& g_{s_2,\ldots,s_{r-1}} := \sum_{j=0}^{\#_+\{s_i\}} {\#_+\{s_i\}\choose j} I_{j + \frac{1}{2}\#_z\{s_i\}} \label{Jbar}
\end{eqnarray} 
where
$\#_s\{s_i\}$ denotes the number of indices in the set $\{s_i\}$ having a value $s$. Here
$I_k := \int_{{\cal D}_m} d^2\varphi f(\varphi)(\cot\varphi)^{2k}$ are elementary integrals which can be evaluated as
\begin{eqnarray}
&& I_k = -\frac{2\pi}{m(2k+1)(\sin \pi/m)^{2k+2}} \sum_{j=0}^{2k+1} {2k+1\choose j} (-1)^j \times
 \nonumber  \\
&& (\cos\pi/m)^{2k+1-j}  \left(
 \mathrm{sinc}(\pi (j+1)/m)-\mathrm{sinc}(\pi(j-1)/m)\right),\nonumber
\end{eqnarray}
and $I_{k+1/2}=0$ for $k$ integer.
The coefficient of Eq. (\ref{Jbar}) $(B^{s_2}\cdots B^{s_{r-1}})_{11}$ is the $(1,1)$-component of a 
product of $(m-1)\times(m-1)$ matrices $B^s$, related to {\em modified Lax operator} \cite{tomaz_quasilocal13},
\begin{eqnarray}
B^0_{j,k} &=& \cos(\pi j l/m) \delta_{j,k}, \quad B^z_{j,k}= -\sin(\pi j l/m) \delta_{j,k}, \\
B^-_{j,k} &=& \sin(\pi k l/m) \delta_{j+1,k}, \quad B^+_{j,k}=-\sin(\pi j l/m) \delta_{j,k+1}. 
\nonumber \label{lax}
\end{eqnarray}
Pauli matrices $\sigma^s_j$ are related to fermion operators via Jordan-Wigner transformation 
$c_j = (\prod_{i=1}^{j-1} \sigma^z_i) \sigma^-_j$. 
The result (\ref{Jbar}) is derived in the limit  $L\to \infty$ and is valid up to corrections of order 
${\cal O}(1/L)$ for a finite periodic ring.
Explicitly, $\bar{J}$ to all terms up to order four ($r\le 4$) reads
\begin{eqnarray}
\bar{J} &=& \tilde{D} \left(8J + 2VJ'\right) + \sum_{j} \Bigl(i \kappa c^\dagger_{j+3}  c_{j} + 
i \kappa'  c^\dagger_{j+3} c^\dagger_{j+2} c_{j+1} c_j \nonumber\\
&&+ i\kappa'' c^\dagger_{j+3} \tilde{n}_{j+2}\tilde{n}_{j+1} c_{j} + {\rm h.c.}\Bigr)+\ldots  \label{Jbarexpl} 
\end{eqnarray}
For example, for $V=1$, ($\Delta=\cos\frac{\pi}{3}$), one has explicitly 
\begin{equation}
\tilde{D}=\frac{1}{8}-\frac{3\sqrt{3}}{32\pi}, \quad
\kappa=\frac{1}{4}-\frac{9\sqrt{3}}{16\pi}, \quad
\kappa'=\frac{9\sqrt{3}}{8\pi}-1, 
\end{equation}
while $\kappa''= \tilde{D} V^2/16$ in general.

\begin{figure}
\includegraphics[width=0.5\textwidth]{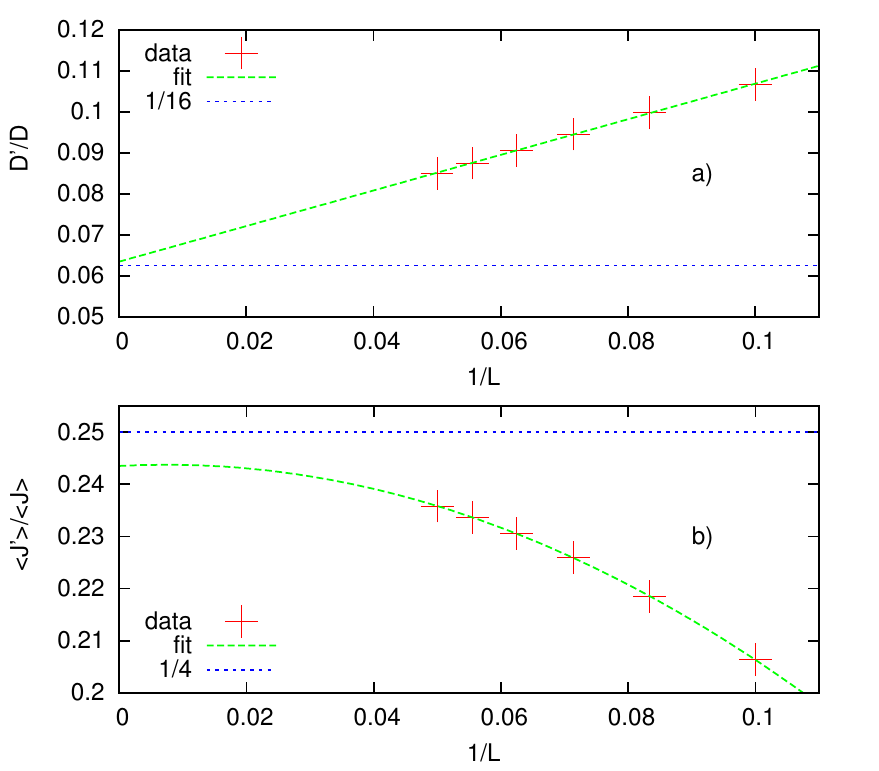}
\caption{ (Color online) a) $D'/D$ vs. $1/L$, where 
$D'$ is the stiffness related with $J'$. b) $\langle J \rangle /\langle J' \rangle$ obtained for 
$\bar{\rho}$, Eq.(\ref{rho2}), for $\beta \to 0$.
Horizontal lines show analytical results. Exact diagonalization has been carried
out for $V=1$ with  $\phi=\pi/L$ and $2 \pi/L$ for even and odd $N$, respectively.}
\label{fig2}
\end{figure}

Above analytical results are nicely corroborated by exact numerical simulations in finite systems 
shown in Fig.~\ref{fig2}. From Eq.~(\ref{Jbarexpl}) one finds that the ratio of two currents 
should be given as ${\rm Tr}\{ \bar{\rho}J' \} /{\rm Tr}\{ \bar{\rho}J \}=V/4$  as confirmed in Fig.~\ref{fig2}b. 
Furthermore, one can define the stiffness with respect to current $J'$ as 
$D'=\ave{\beta \overline{J'}^2}/(2L)$. 
Formula (\ref{Jbarexpl}) immediately implies that $\overline{J'} = (V/4) \bar{J}$, and so
the two stiffnesses should have a simple ratio $D'/D = (V/4)^2$ (see Fig.~\ref{fig2}a). 

In conclusion, we have proposed a class of global quantum quench dynamics of integrable spin chains for 
which the state at asymptotic times does not relax to GGE. We argue that, at least for weak quenches 
where linear response theory is applicable, the validity of GGE ensemble is in one-to-one correspondence 
with the saturation of the Mazur bound expressed in terms of strictly local conserved operators. However, 
if one extends the GGE ensemble by including the quasi-local conserved operators from the opposite parity sector 
-- having linearly  extensive Hilbert-Schmidt norm -- then the latter can be used to describe exactly the steady 
state density 
operator after the quench. Our theory has been demonstrated in the 1D model of interacting spinless
fermions (XXZ spin model) within the metallic regime.

It should be noted that our results are expected to have further implications on other relevant quantities 
of integrable system 
besides the charge stiffness. The flux--quench induced steady current 
$\langle \bar{J} \rangle = 2 \sum_k \sin(k) \langle \bar{n}_k \rangle  \ne 0$ 
is reflected
into the fermion momentum--distribution function $\langle \bar{n}_k \rangle$ which also does not comply 
to the standard GGE.
The latter quantity is the one typically 
measured in cold--atom experiments \cite{bloch1,bloch2} as well most frequently studied in connection 
with the GGE concept \cite{Manmana2007,gge,Cassidy2011}. 
The inclusion of the quasi--local conserved quantity $\bar J$ fully fixes the   
steady state $\langle \bar{n}_k \rangle$ within our quench protocol via extended GGE form Eq.~(\ref{rho2}). 
It is still tempting to construct and consider further 
(presumably conserved) quantities from the same polarity sector which would fix this
and related quantities for an arbitrary quench. 

\acknowledgements
M.M. acknowledges support from the NCN project  DEC-2013/09/B/ST3/01659. P.P. and T.P. acknowledges 
the support by the program P1-0044 and projects J1-4244 (P.~P.) and J1-5349 (T.~P.) of the Slovenian Research Agency.

\bibliography{bib_nongge.bib}

\end{document}